\documentstyle[12pt,psfig]{article}

\def\TL{\hfil$\displaystyle{##}$}
\def\TR{$\displaystyle{{}##}$\hfil}

\def\comment#1{}
\def\fixit#1{}

\def\tf#1#2{{\textstyle{#1 \over #2}}}

\def\mop#1{\mathop{\rm #1}\nolimits}

\def\Vol{\mop{Vol}}
\def\vol{\mop{vol}}

\def\tr{\mop{tr}}

\def\lsim{\mathrel{\mathstrut\smash{\ooalign{\raise2.5pt\hbox{$<$}\cr\lower2.5pt\hbox{$\sim$}}}}}
\def\gsim{\mathrel{\mathstrut\smash{\ooalign{\raise2.5pt\hbox{$>$}\cr\lower2.5pt\hbox{$\sim$}}}}}

\def\sqr#1#2{{\vcenter{\vbox{\hrule height.#2pt
         \hbox{\vrule width.#2pt height#1pt \kern#1pt
            \vrule width.#2pt}
         \hrule height.#2pt}}}}
\def\square{\mathop{\mathchoice\sqr56\sqr56\sqr{3.75}4\sqr34}\nolimits}

\def\href#1#2{#2}  

\def\lbldef#1#2{\expandafter\gdef\csname #1\endcsname {#2}}
\def\eqn#1#2{\lbldef{#1}{(\ref{#1})}%
\begin{equation} #2 \label{#1} \end{equation}}
\def\eqalign#1{\vcenter{\openup1\jot
    \halign{\strut\span\TL & \span\TR\cr #1 \cr
   }}}
\def\eno#1{(\ref{#1})}

\begin{document}
\begin{titlepage}

\begin{flushright}
PUPT-1805 \\
hep-th/9807164
\end{flushright}
\vfil\vfil

\begin{center}

{\Large {\bf Einstein Manifolds and Conformal Field Theories}}

\vfil

Steven S.~Gubser

\vfil

Joseph Henry Laboratories\\
 Princeton University\\
 Princeton, New Jersey 08544\\

\vfil

\end{center}


\vspace{5mm}

\begin{abstract} 
\noindent 
 In light of the AdS/CFT correspondence, it is natural to try to
define a conformal field theory in a large $N$, strong coupling limit
via a supergravity compactification on the product of an Einstein
manifold and anti-de Sitter space.  We consider the five-dimensional
manifolds $T^{pq}$ which are coset spaces $(SU(2) \times SU(2))/U(1)$.
The central charge and a part of the chiral spectrum are calculated,
respectively, from the volume of $T^{pq}$ and the spectrum of the
scalar laplacian.  Of the manifolds considered, only $T^{11}$ admits
any supersymmetry: it is this manifold which characterizes the
supergravity solution corresponding to a large number of D3-branes at
a conifold singularity, discussed recently in \cite{kw}.  Through a
field theory analysis of anomalous three point functions we are able
to reproduce the central charge predicted for the $T^{11}$ theory by
supergravity: it is $27/32$ of the central charge of the ${\cal N}=2$
${\bf Z}_2$ orbifold theory from which it descends via an RG flow.

\end{abstract}

\vfil\vfil\vfil
\begin{flushleft}
July 1998
\end{flushleft}
\end{titlepage}

\newpage
\renewcommand{\baselinestretch}{1}   
\baselineskip14.5pt                  
\section{Introduction}
\label{Introduction}

A large number of coincident D3-branes in a smooth spacetime induce a
supergravity metric which is locally $AdS_5 \times S^5$.  It is in the
context of this geometry that the AdS/CFT correspondence was first
developed \cite{JuanAdS97,gkPol,WitHolOne}.  Considerable thought has
already been given to D3-branes on orbifold singularities
\cite{ksOrb,lnv}, as components of F-theory vacua \cite{Kehag}, in the
presence of D7-branes and orientifolds \cite{fsOrient,kz,afm} and more
recently on conifold singularities \cite{kw}.  In each case the
geometry can be worked out, and properties of the D3-brane
world-volume theory can be deduced from the holographic prescriptions
of \cite{gkPol,WitHolOne}.

An alternative approach, which we take in this paper, is to start with
a geometry and see how its properties relate via the holographic
correspondence to conformal field theory.  The conjecture is implicit
in \cite{WitHolOne} that {\it any} anti-de Sitter vacuum of string
theory or M-theory defines a conformal field theory.\footnote{Since
one needs a stress tensor to define a CFT and hence a graviton in the
bulk spectrum, it seems that a theory containing gravity is necessary
in the bulk to make the conjecture reasonable.  If anything more than
a large $N$ limit of the boundary theory is desired, the bulk theory
must be quantum mechanical.  It is for this reason that we consider
string theory or M-theory the only candidates for the bulk theory.}
The simplest examples apart from $S^5$ involve the coset manifolds
$T^{pq}$ considered in \cite{romans}.  Of these, only $T^{11}$
preserves some supersymmetry (${\cal N}=1$, which is $1/4$ of maximal
considering conformal invariance).  One is entitled to wonder what
status the other $T^{pq}$ compactifications have as string vacua.  In
any case, whatever exotic quantum field theoretic behavior reflects
the usual pathologies of non-supersymmetric string vacua, it is at
least a fascinating consequence of holography that a conformal fixed
point at large $N$ and strong coupling exists and is characterized by
compactified classical type IIB supergravity.

In section~\ref{TpqVol} we use the volume of the Einstein manifold to
compute the central charge of the conformal field theory.  The results
tally with field theory in the case of orbifolds, but for the coset
manifolds the central charge is typically irrational, which rules out
any weakly interacting lagrangian description.  For the case of
$T^{11}$ supported by $N$ units of D3-brane charge, the central charge
is $27/16$ times the central charge for ${\cal N}=4$ $U(N)$
super-Yang-Mills theory.  Through an analysis, presented in
section~\ref{FTAnom}, of anomalous three point functions of the
$R$-current and the stress-energy tensor, we are able to reproduce
this number on the field theory side.

In section~\ref{TpqLap} we compute the spectrum of the scalar
laplacian on the $T^{pq}$ manifolds.  Typically, dimensions of
operators are irrational as well.  In the case of $T^{11}$, only part
of this spectrum can be related to chiral primaries and their
descendants; most of the others again have irrational dimensions.
There is no reason to suppose that these dimensions do not flow as
couplings are changed, but this makes their finite values in the limit
of supergravity's validity all the more remarkable: they provide an
example of operators whose dimensions are not protected but
nevertheless tend to finite values in the strong coupling limit.

\section{Volume and curvature of $T^{pq}$}
\label{TpqVol}

From the basic AdS/CFT setup as enunciated in \cite{gkPol,WitHolOne},
one can conclude that the central charge of the conformal field theory
is inversely proportional (in the large $N$ limit at least) to the
volume of the compact five-dimensional Einstein manifold $M_5$.  To
see this, consider the geometry
  \eqn{EinGeom}{\eqalign{
   ds^2 &= {L^2 \over z^2} (d\vec{x}^2 + dz^2) + L^2 ds_{M_5}^2  \cr
   F &= {N\sqrt{\pi} \over 2 \Vol M_5} 
    \left( \vol_{M_5} + \vol_{AdS} \right)
  }}
 where the metric $ds_{M_5}^2$ is taken to have curvature
$R_\alpha{}^\beta = 4 \delta_\alpha^\beta$.  The volume $\Vol M_5$ and
the five-forms $\vol_{M_5}$ and $\vol_{AdS}$ refer to the metrics
$ds_{M_5}^2$ and $\tf{1}{z^2} (d\vec{x}^2 + dz^2)$ (ie, without the
powers of $L$ which appear in the full ten-dimensional metric).  $N$
is the integer 3-brane charge.  The Einstein equation,
  \eqn{EinEq}{
   R_M{}^N = {\kappa^2 \over 6} F_{M P_1 P_2 P_3 P_4} 
    F^{N P_1 P_2 P_3 P_4} \ ,
  }
 is satisfied if one takes 
  \eqn{LReq}{
   L^4 = {\sqrt{\pi} \over 2} {\kappa N \over \Vol M_5} \ .
  }
 After compactification on $M_5$, type~IIB supergravity can be
summarized by the action
  \eqn{IIBAct}{
   S = {\Vol M_5 \over 2 \kappa^2} L^8 \int d^5 x \, \sqrt{g}
    \left( R + 12 - \tf{1}{2} (\partial\phi)^2 + \ldots \right) \ .
  }
 In \IIBAct, the metric is taken to be $ds^2 = g_{\mu\nu} dx^\mu
dx^\nu = \tf{1}{z^2} (d\vec{x}^2 + dz^2)$: we have rescaled out all
factors of $L$ from the integrand.  In view of \LReq, the prefactor on
$S$ in \IIBAct\ is ${\pi N^2 \over 8 \Vol M_5}$.  All correlators
calculated via the holographic prescription of \cite{gkPol,WitHolOne}
include an overall factor $1/ \Vol M_5$ from this prefactor. 

Indeed, one of the first checks made in \cite{gkPol} was to see that
the central charge of the ${\cal N} = 4$ theory came out correctly
from the two-point function of the stress-energy tensor, calculated
essentially via \IIBAct\ with $M_5 = S^5$.  In a more involved
calculation, the authors of \cite{hsWeyl} succeeded in establishing
  \eqn{OneWeyl}{
   \langle T^\alpha{}_\alpha \rangle_{g_{\mu\nu}} = 
    -a E_4 - c I_4
  }
 where
  \eqn{EIWeyl}{\eqalign{
   E_4 &= {1 \over 16 \pi^2} \left( R_{ijkl}^2 - 4 R_{ij}^2 + 
     R^2 \right)  \cr
   I_4 &= -{1 \over 16 \pi^2} \left( R_{ijkl}^2 - 2 R_{ij}^2 +
     \tf{1}{3} R^2 \right)
  }}
 from the holographic prescription applied to an arbitrary boundary
metric $g_{\mu\nu}$ in the conformal class of (compactified) Minkowski
space.  The calculation again follows from \IIBAct\ (improved by
boundary terms and regulated), so $a$ and $c$ carry an inverse factor
of $\Vol M_5$.  The coefficient $c$ is the same as what we have called
the central charge (namely the normalization of the two-point function
of the stress-energy tensor), so the first term of \OneWeyl\ follows
from what was already worked out in \cite{gkThree}.  The derivation of
\cite{hsWeyl} shows that $a=c$ is a consequence of holography plus the
product space structure of the spacetime (thus restricting the types
of theories one could hope to construct holographically from product
geometries).  It was also reported there that the agreement between
the holographic and free field values of $a$ and $c$ persists to
orbifolded theories such as those considered in \cite{ksOrb}.  The
free field counting is done using the following table \cite{DuffWeyl}
(all other entries follow from the first three):
  \eqn{CCC}{
   \begin{tabular}{|l||c|c|} \hline
    field                          &$ a      $&$ c     $ \\ \hline
    real scalar                    &$ 1/360  $&$ 1/120 $ \\
    complex Weyl fermion           &$ 11/720 $&$ 1/40  $ \\
    vector boson                   &$ 31/180 $&$ 1/10  $ \\
    ${\cal N}=1$ chiral multiplet  &$ 1/48   $&$ 1/24  $ \\
    ${\cal N}=1$ vector multiplet  &$ 3/16   $&$ 1/8   $ \\
    ${\cal N}=2$ hyper multiplet   &$ 1/24   $&$ 1/12  $ \\
    ${\cal N}=2$ vector multiplet  &$ 5/24   $&$ 1/6   $ \\
    ${\cal N}=4$ vector multiplet  &$ 1/4    $&$ 1/4 $   \\ \hline
   \end{tabular}
  }
 Counting the free field content is a reliable method for computing
the anomaly coefficients as long as there is at least ${\cal N}=1$
supersymmetry, since then $T^\mu{}_\mu$ and $\partial_\mu R^\mu$ are
superpartners and the Adler-Bardeen theorem may be applied.  In fact,
the results of \cite{bkv,bj} imply that the free field counting is
valid to leading order in large $N$ even for non-supersymmetric
theories obtained by orbifolding the ${\cal N}=4$ theory.  Indeed, all
the agreements obtained so far are strictly leading large $N$ limits:
the exact result $a=c=(N^2-1)/4$ for ${\cal N}=4$ $SU(N)$ gauge theory
apparently includes a one-loop contribution from supergravity.

To learn the central charge of the $T^{pq}$ theories, we must compute
the volume of these manifolds.  We will use the methods of \cite{CRW}.
There are a few important differences between the conventions used
here and those of \cite{CRW} and \cite{romans}, and we will write out
formulas explicitly enough to disambiguate them.  To form the quotient
space $T^{pq}$ we start with $SU(2) \times SU(2)$ generated by $i
\sigma_k$ and $i \tau_k$, where $k$ runs from $1$ to $3$, and divide
out by the $U(1)$ generated by $\omega = p \sigma_3 + q \tau_3$.  Let
us write the generators as
  \eqn{RotGen}{
   i \sigma_l \ , \quad i \tau_s \ , \quad
   Z = q i \sigma_3 - p i \tau_3 \ , \quad
   \omega = p i \sigma_3 + q i \tau_3 
  }
 where $l$ and $s$ run over $1,2$.  Referring to these anti-hermitian
generators collectively as $t_a$, we define the structure constants and
Killing metric so that
  \eqn{Cgamma}{\eqalign{
   [t_a,t_b] &= C_{ab}^c t_c  \cr
   \gamma_{ab} &= -\tf{1}{8} C_{ac}^d C_{bd}^c \ .
  }}
 The normalization of $\gamma_{ab}$ is chosen so that this is the same
metric as the natural one on $S^3 \times S^3$ where $S^3$ is the unit
three-sphere.  In this metric, the volume of $SU(2) \times SU(2)$ is
$(2 \pi^2)^2$, and the length of the orbit of $\omega$ is $2 \pi
\sqrt{p^2+q^2}$.  So before the symmetric rescaling, 
  \eqn{VolBefore}{
   \Vol T^{pq} = {2 \pi^3 \over \sqrt{p^2+q^2}} \ .
  }
 The symmetric rescaling referred to in \cite{romans} and detailed in
\cite{CRW} is a replacement $e^a \to r(a) e^a$, where $e^a$ is the
vielbein for the coset space generated by $i\sigma_l$, $i\tau_s$, and
$Z$.  In a (hopefully) obvious notation, let us choose
  \eqn{raDef}{
   r(l) = \sqrt{a} \gamma \quad r(s) = \sqrt{b} \gamma \quad 
    r(Z) = \gamma \ .
  }
 The volume in the new metric, which we shall call the squashed
metric, is
  \eqn{VolAfter}{
   \Vol T^{pq} = {2 \pi^3 \over \sqrt{p^2+q^2}} {1 \over a b \gamma^5} \ .
  }
 The squashed metric still has an isometry group $SU(2) \times SU(2)
\times U(1)$.  Furthermore, with a special choice of $a$ and $b$, the
metric can be made Einstein.  In \cite{CRW} the following expression
is given for the Riemann tensor (up to a convention-dependent factor
of $2$):\footnote{In \eno{RCDerive} the Riemann tensor has been
defined from the spin connection via $R^a{}_b = \tf{1}{2}
R^a{}_{b\mu\nu} dy^\mu dy^\nu = d\omega + \omega \wedge \omega$,
whereas in \cite{CRW} the factor of $\tf{1}{2}$ is omitted.  In
\cite{romans} the normalization seems to be the same as used here.}
  \eqn{RCDerive}{\eqalign{
   R^a{}_{bcd} &= \tf{1}{2} C_{bc}^a C_{de}^c 
     \left( {a\ b \atop c} \right) + 
    C_{b\omega}^a C_{de}^\omega r(d) r(e)  \cr &\quad + 
    \tf{1}{4} C_{cd}^a C_{be}^c \left( {a\ c \atop d} \right)
     \left( {b\ c \atop e} \right) - 
    \tf{1}{4} C_{ce}^a C_{bd}^c \left( {a\ c \atop e} \right)
     \left( {b\ c \atop d} \right)  \cr 
   \left( {a\ b \atop c} \right) &\equiv 
    {r(a) r(c) \over r(b)} + {r(b) r(c) \over r(a)} -
     {r(a) r(b) \over r(c)} \ .
  }}
 From the Einstein requirement $R_a{}^b = \Lambda \delta_a^b$ one now
derives equation (2.5) of \cite{romans}:
  \eqn{TwoFive}{
   {\Lambda \over \gamma^2} = 
    4 a - 2 a^2 y^2 = 4 b - 2 b^2 x^2 = 2 a^2 y^2 + 2 b^2 x^2
  }
 where $x = p / \sqrt{p^2+q^2}$ and $y = q / \sqrt{p^2+q^2}$.  Note
that for given $x$ and $y$, the determination of $a$ and $b$ reduces
to solving a cubic.

To sum up the discussion at the beginning of this section, the ratio
of the central charge $c$ in the conformal field theory ``defined'' by
an Einstein manifold $M_5$ to the central charge $c_0 = N^2/4$ of
${\cal N} = 4$ super-Yang-Mills with gauge group $U(N)$ should be the
same as the ratio of $\Vol S^5$ to $\Vol M_5$ computed with respect to
Einstein metrics with the same cosmological constant.  Using \TwoFive\
and the fact that the unit $S^5$ has $\Vol S^5 = \pi^3$ and $\Lambda =
4$, one arrives at
  \eqn{cRatio}{
   {c \over c_0} = 2\sqrt{2} {a b (p^2 + q^2)^3 \over 
     (q^2 a^2 + p^2 b^2)^{5/2}} \ .
  }
 The results for the three simplest examples are quoted in the following
table:
  \eqn{TableABC}{
   \begin{tabular}{|c||c|c|c|c|c|c|} \hline
    $T^{pq}  $&$ x $&$ y $&$ a $&$ b $&$ \gamma $&$ c/c_0 $ \\ \hline
    $T^{01}  $&$ 0 $&$ 1 $&$ 1 $&$ 1/2 $&$ \sqrt{2} $&$ \sqrt{2} $ \\
    $T^{11}  $&$ 1/\sqrt{2} $&$ 1/\sqrt{2} $&$ 4/3 $&$ 4/3 $&$ 
     3/\sqrt{8} $&$ 27/16 $ \\ 
    $T^{34}  $&$ 3/5 $&$ 4/5 $&$ 25/18 $&$ 25/27 $&$ 
     \tf{9}{5} \sqrt{\tf{2}{5}} $&$ 
     \tf{243}{25} \sqrt{\tf{2}{5}} $ \\ \hline
   \end{tabular}
  }
 The value listed for $\gamma$ is the one which makes $\Lambda = 4$.
This normalization will be useful in the next section.

The higher $T^{pq}$ have very complicated $a$ and $b$, and it does not
seem worth the space to quote their central charges; suffice it to say
that $c/c_0$ is typically an irrational involving irreducible square
and cube roots.  One can show using \TwoFive\ and \cRatio\ that $c/c_0
\geq O(\sqrt{pq})$ (a better bound may be possible).

\subsection{The anomaly coefficients in field theory}
\label{FTAnom}

In view of the construction \cite{kw} of the $T^{11}$ theory from a
relevant deformation of the ${\cal N}=2$ ${\bf Z}_2$ orbifold theory,
one would expect to be able to derive the central charge in \TableABC\
from a field theory analysis along the lines of \cite{afgj}.  This is
quite a nontrivial test of holography because it tests a
nonperturbative field theoretic effect involving an RG flow from a
simple UV theory to a IR theory which lies deep inside the conformal
window. (Because $N_f = 2N_c$ for each of the gauge groups separately,
the IR theory is not close to either edge of the conformal window
\cite{sei} $3N_c/2 \leq N_f \leq 3N_c$.  Perturbation theory in the
electric or magnetic representation, respectively, can be used if $N_f
= 3 N_c - \epsilon$ or $N_f = 3 N_c/2 + \epsilon$).  In an ${\cal
N}=1$ superconformal theory, the anomaly coefficients $a$ and $c$ can
be read off from anomalous three point functions $\partial_\mu \langle
T_{\alpha\beta} T_{\gamma\delta} R^\mu \rangle$ and $\partial_\mu
\langle R_\alpha R_\beta R^\mu \rangle$, where $R_\mu$ is the
$R$-current: the former is proportional to $a-c$, while the latter is
proportional to $5a-3c$.

The ultraviolet theory, in ${\cal N}=1$ language, has $2N_c^2$ vector
multiplets filling out the adjoints of the two $U(N_c)$ gauge groups,
plus $6N_c^2$ chiral multiplets filling out an adjoint for each gauge
group ($2N_c^2$) plus bifundamental matter ($4N_c^2$).  The
$R$-current descends from one of the ${\cal N}=4$ $SU(4)$
$R$-currents.  The $R$-charge of the $N_\lambda = 2N_c^2$ gluino
fields is $1$, whereas the $R$-charge of the $N_\chi = 6N_c^2$ quark
fields is $-1/3$.  There is no gravitational anomaly in $\langle T T R
\rangle$ because the $U(1)_R$ generator is traceless.  So
$a_{UV}-c_{UV}=0$.  For the $\partial_\mu \langle R_\alpha R_\beta
R^\mu \rangle$ anomaly, we use the position space analysis summarized
in \cite{afgj}: for a single Majorana spinor with axial current $J_\mu
= \tf{1}{2} \bar\psi \gamma_\mu \gamma^5 \psi$,\footnote{There is a
small typo in (4.10) of \cite{afgj} which is corrected in
\eno{JAnom}.}
  \eqn{JAnom}{
   {\partial \over \partial z^\mu} 
    \langle J_\alpha(x) J_\beta(y) J^\mu(z) \rangle = 
    -{1 \over 12\pi^2} \epsilon_{\alpha\beta\gamma\delta}
      {\partial \over \partial x^\gamma} {\partial \over \partial y^\delta}
      \delta^4(x-z) \delta^4(y-z) \equiv 
      \tf{9}{16} {\cal A}_{\alpha\beta}(x,y,z) \ .
  }
 By counting $U(1)$ charges, one obtains 
  \eqn{RAnom}{
   \partial_\mu \langle R_\alpha R_\beta R^\mu \rangle = 
    (5a_{UV}-3c_{UV}) {\cal A}_{\alpha\beta} = 
    \tf{9}{16} \left( N_\lambda + 
     \left(-\tf{1}{3}\right)^3 N_\chi \right) {\cal A}_{\alpha\beta} =
    N_c^2 {\cal A}_{\alpha\beta} \ .
  }
 So far, this has just verified the known result from free field
counting with \CCC\ that $a_{UV}=c_{UV}=N_c^2/2$.

The relevant deformation discussed in \cite{kw} gives a mass to the
chiral multiplets in the adjoints of the gauge groups, leaving behind
$4 N_c^2$ chiral multiplets with a quartic superpotential, plus $2
N_c^2$ vector multiplets as before.  The anomalous dimension of mass
operators of the form $\tr AB$ was found in the infrared \cite{kw}
from the exact beta function \cite{ShVOne,ShVTwo} to be $\gamma_{IR} =
-1/2$.  After the adjoint chirals have been made massive and
integrated out, the $R$-current which is the superpartner of the
stress-energy tensor is no longer conserved, in part due to internal
anomalies proportional to the beta function.  A non-anomalous $U(1)$
current \cite{ksv} is $S_\mu = R_\mu + \tf{1}{3} (\gamma_{IR} -
\gamma) K_\mu$, where $K_\mu$ is the Konishi current, which assigns
charge $1$ to quarks and $0$ to gluinos.\footnote{S.~P.~de Alwis has
shown \cite{deal} how one can define the current $S_\mu$ in the full
${\cal N}=2$ ${\bf Z_2}$ orbifold theory instead of first integrating
out the chiral adjoint fields, as we have done here.  \eno{SAnom} is
unchanged because the $S$-charge of the fermions in those chiral
adjoints is $0$ in the ultraviolet.}
 The external gauge and gravitational anomalies of $S_\mu$ are
independent of scale.  Since $S_\mu = R_\mu$ at the IR fixed point,
the strategy is to evaluate the external anomalies of $S_\mu$ at a
high scale where $\gamma \to 0$, and identify the answers as the IR
fixed point anomalies of $R_\mu$.  When $\gamma \to 0$, the
$S$-charges of the fermion fields which remain massless are $1$ for
the $N_\lambda = 2 N_c^2$ gluinos and $-1/2$ for the $N_\chi = 4
N_c^2$ quarks.  As before, there is no gravitational anomaly because
the $U(1)_S$ generator is traceless.  So $a_{IR}-c_{IR}=0$.  On the
other hand,
  \eqn{SAnom}{
   \partial_\mu \langle R_\alpha R_\beta R^\mu \rangle = 
    (5a_{IR}-3c_{IR}) {\cal A}_{\alpha\beta} = 
    \tf{9}{16} \left( N_\lambda + 
     \left(-\tf{1}{2}\right)^3 N_\chi \right) {\cal A}_{\alpha\beta} =
    \tf{27}{32} N_c^2 {\cal A}_{\alpha\beta} \ .
  }
 Thus $a_{IR}/a_{UV} = c_{IR}/c_{UV} = 27/32$, exactly the ratio of
$\Vol S^5/{\bf Z}_2$ to $\Vol T^{11}$.  Of course, the agreement with
supergravity is only to leading order in large $N$.  A more careful
analysis of the field theory would have the $U(1)$ parts of the gauge
groups decoupling, changing $a$ and $c$ by a factor $1-O(1/N^2)$ which
is invisible in classical supergravity.

\section{The laplacian on $T^{pq}$}
\label{TpqLap}

To study the spectrum of chiral primaries in the $T^{pq}$ CFT (insofar
as we grant that the theory exists and is defined in the large $N$
limit by the supergravity), the first step is to find the spectrum of
the scalar laplacian on $T^{pq}$.  A convenient parametrization of
these coset spaces descends from Euler angles on $SU(2) \times SU(2)$.
Let's write a group element of $SU(2)$ as $g = e^{i \alpha J_z} e^{i
\beta J_y} e^{i \gamma J_z}$.  The standard unit sphere metric on
$SU(2)$ is, in these variables,
  \eqn{SUTwoMet}{
   ds^2 = \tf{1}{4} \left( d\alpha^2 + d\beta^2 + d\gamma^2 + 
    2 \cos\beta \, d\alpha d\gamma \right) \ .
  }
 The angles are allowed to vary over the ranges
  \eqn{EulerRanges}{
   \alpha \in (0,2\pi) \quad
   \beta \in (0,\pi) \quad
   \gamma \in (0,4\pi) \ .
  }
 The reason that $\alpha$ is not allowed to range over $(0,4\pi)$ is
to avoid a double cover of $SU(2)$: $e^{2 \pi i J_z} = -1$.

 The convenience of Euler angles is that we can represent the coset
space $T^{pq}$ as the hypersurface $p \gamma_1 + q \gamma_2 = 0$ in
the Euler angle coordinatization
$(\alpha_1,\beta_1,\gamma_1,\alpha_2,\beta_2,\gamma_2)$ of $SU(2)
\times SU(2)$.  Writing $\gamma_1 = -q \alpha_3 / \sqrt{p^2+q^2}$,
$\gamma_2 = p \alpha_3 / \sqrt{p^2+q^2}$, and ordering our coordinates
for $T^{pq}$ as $(\beta_1,\beta_2,\alpha_1,\alpha_2,\alpha_3)$, one
arrives at a metric whose inverse is 
  \eqn{ExSqMet}{
   g^{ab} = 4 \gamma^2 
   \pmatrix{ a & 0 & 0 & 0 & 0 \cr
             0 & b & 0 & 0 & 0 \cr
             0 & 0 & a \csc^2 \beta_1 & 0 & y a \cot \beta_1 \csc \beta_1 \cr
             0 & 0 & 0 & b \csc^2 \beta_2 & -x b \cot \beta_2 \csc \beta_2 \cr
             0 & 0 & y a \cot \beta_1 \csc \beta_1 &
              -x b \cot \beta_2 \csc \beta_2 &
              1 + y^2 a \cot^2 \beta_1 + x^2 b \cot^2 \beta_2}
  }
 after squashing.  One can now verify \VolAfter\ by integrating
$\sqrt{g} = \sqrt{\det g_{ab}} = ab\gamma^5 \sin\beta_1 \sin\beta_2$
over the allowed ranges for the variables:
  \eqn{VarRange}{
   \beta_1,\beta_2 \in (0,\pi) \qquad
   \alpha_1,\alpha_2 \in (0,2\pi) \qquad
   \alpha_3 \in \left( 0,4\pi/\sqrt{p^2+q^2} \right) \ .
  }

The laplacian has a fairly simple form:
  \eqn{LapEq}{\eqalign{
   \square \phi &= 4 \gamma^2 \Bigg[ a {1 \over \sin\beta_1} 
     {\partial \over \partial \beta_1}
     \sin\beta_1 {\partial \over \partial \beta_1} + 
     b {1 \over \sin\beta_2} 
     {\partial \over \partial \beta_2}
     \sin\beta_2 {\partial \over \partial \beta_2}  \cr &\quad + 
      a \csc^2 \beta_1 {\partial^2 \over \partial \alpha_1^2} + 
      b \csc^2 \beta_2 {\partial^2 \over \partial \alpha_2^2}  \cr &\quad + 
      2ya \cot\beta_1 \csc\beta_1 
       {\partial^2 \over \partial\alpha_1 \partial\alpha_3} -
      2xb \cot\beta_2 \csc\beta_2
       {\partial^2 \over \partial\alpha_2 \partial\alpha_3}  \cr &\quad + 
      (1 + y^2 a \cot^2 \beta_1 + x^2 b \cot^2 \beta_2) 
       {\partial^2 \over \partial\alpha_3^2} \Bigg] \phi = -E \phi \ .
  }}
 Amusingly enough, this equation can be solved completely by
separation of variables even though $T^{pq}$ is not metrically a
product space: writing
  \eqn{SepAnsatz}{
   \phi = \phi_1(\beta_1) \phi_2(\beta_2) 
    \exp\left( i \sum_{j=1}^3 m_j \alpha_j \right)
  }
 we arrive at
  \eqn{ESum}{
   E = 4 \gamma^2 (a E_1 + b E_2 + m_3^2)
  }
 where $E_1$ and $E_2$ are determined by the ordinary differential
equations 
  \eqn{ODEi}{
   \left[ {1 \over \sin\beta_i} {\partial \over \partial \beta_i}
    \sin\beta_i {\partial \over \partial \beta_i} - 
    \left( m_3 y_i \cot\beta_i + m_i \csc\beta_i \right)^2 \right]
    \phi_i = -E_i \phi_i
  }
 where $y_1 = y$ and $y_2 = -x$.  For $y_i = 0$ and $1$, we recognize
\ODEi\ as the differential equation that determines eigenvalues of the
laplacian on $S^2$ and $S^3$, respectively: in these cases one would
have $E_i = l(l+1)$ or $l(l+2)/4$.

The next step is to determine, given specified $\eta$ and $\chi$, the
values of $E$ for which
  \eqn{ODEarch}{
   \left[ {1 \over \sin\beta} {\partial \over \partial \beta}
    \sin\beta {\partial \over \partial \beta} - 
    \left( \eta \cot\beta + \chi \csc\beta \right)^2 \right]
    \phi = -e \phi
  } 
 admits a regular solution on the interval $\beta \in [0,\pi]$.
Introducing a new variable $z = \cos^2 {\beta \over 2}$, one can
reduce \ODEarch\ to a hypergeometric equation.  Solutions of the form
  \eqn{HyperForm}{
   \phi = z^F (1-z)^G F(A,B;C;z)
  }
 with suitable $A$, $B$, $C$, $F$, and $G$ are smooth in the interior
of the interval, and have a behavior at the endpoints which can be
determined using the formula \cite{GR}
  \eqn{HyperTrans}{\eqalign{
   &F(A,B;C;z) = {\Gamma(C) \Gamma(C-A-B) \over \Gamma(C-A) \Gamma(C-B)}
     F(A,B;A+B-C+1;1-z)  \cr &\qquad + 
    (1-z)^{C-A-B} {\Gamma(C) \Gamma(A+B-C) \over \Gamma(A) \Gamma(B)}
     F(C-A,C-B;C-A-B+1;1-z) \ .
  }}
 In brief, the solutions are regular when they can be expressed in
terms of a hypergeometric function which is a polynomial.  This is so
when 
  \def\max{\mop{max}}
  \eqn{MaxSum}{
   \tf{1}{2} - \sqrt{\tf{1}{4} + e + \eta^2} + \max\{|\eta|,|\chi|\}
    \in {\bf Z}^- \equiv \{0,-1,-2,-3,\ldots\} \ .
  }
 One can verify \MaxSum\ in four cases, taking $\eta+\chi$ and
$\eta-\chi$ positive or negative.  The eigenfunctions $\phi$ in fact
vanish at the endpoints except when $|\eta|=|\chi|$.  The case
$\eta=\chi=0$ leads to the Legendre polynomials.  Let us go through
the derivation of \MaxSum\ for the case $\eta-\chi \leq 0$ and
$\eta+\chi \geq 0$, and leave the other cases as an exercise for the
reader.  Choosing $F = -{\eta-\chi \over 2}$ and $G = {\eta+\chi \over
2}$, we are led to
  $$\displaylines{
   A = \tf{1}{2} + \chi - \sqrt{\tf{1}{4} + e + \eta^2} \qquad
   B = \tf{1}{2} + \chi + \sqrt{\tf{1}{4} + e + \eta^2} \cr
   C = 1 + \chi - \eta \ .
  }$$
 Regularity at $z=0$ is guaranteed by our choice of the sign on $F$.
Because of the choice of sign on $G$, regularity at $z=1$ depends on
having $A$ or $B$ vanish, so that the second term in \HyperTrans\ is
absent.  This occurs when $A \in {\bf Z}^-$ or $B \in {\bf Z}^-$.  The
latter is impossible because $\chi \geq |\eta|$; so we are left with
$A \in {\bf Z}^-$, which indeed reduces to \MaxSum.

Writing $l = k + \max\{|\eta|,|\chi|\}$ where $k \in {\bf Z}^+ \equiv
\{0,1,2,3,\ldots\}$, one finds the simple expression $e = l(l+1) -
\eta^2$.  Returning to \ESum, the final expression for the eigenvalues
of the laplacian on $T^{pq}$ is
  \eqn{EForm}{
   E = 4\gamma^2 \left[ a l_1 (l_1+1) + b l_2 (l_2+1) +
    m_3^2 (1-ay^2-bx^2) \right] \ .
  }
 In \EForm, $l_i = k_i + \max\{|m_3 y_i|,|m_i|\}$, $k_i \in {\bf
Z}^+$, $m_i \in {\bf Z} - m_3 y_i$, $m_3 \in {\sqrt{p^2+q^2} \over 2}
{\bf Z}$, and $i=1,2$.  The shift in the allowed values of $m_i$ is
necessary to ensure that \SepAnsatz\ is single valued.  Note that $m_3
y_i$ is always either integer or half-integer, and that $E$ is
completely specified by $m_3$ (essentially, the $U(1)$ charge) and the
spins $l_i$ under the two $SU(2)$'s.  The expression of \EForm\ as a
linear combination of the quadratic casimirs for the symmetry group
$SU(2) \times SU(2) \times U(1)$ is precisely the form expected for a
coset manifold.\footnote{Thanks to E.~Witten for a discussion on this
point.}

As in the case of $S^5$, the dimension of the scalar operator in the
conformal field theory to which a given mode of the dilaton $\phi$
couples is 
  \eqn{DimEn}{
   \Delta = 2 + \sqrt{4 + E} \ ,
  } 
 where now $\gamma$ must be chosen in \EForm\ so that $\Lambda = 4$.
In fact, these modes may be complexified by replacing $\phi$ by a
complex scalar $B$ which also includes the axion.  Comparing (2.33)
and (2.53) of \cite{KRvN}, one arrives at the conclusion that the
operators coupling to the modes of $h^\alpha_\alpha$ and
$a_{\alpha\beta\gamma\delta}$ (which can be mixed together in two
different ways) have dimensions
  \eqn{DimEnTwo}{
   \Delta = \left\{ \eqalign{ & -2 + \sqrt{4 + E}  \cr  
                              &  6 + \sqrt{4 + E} \ .} \right.
  }
 These dimensions can be read off directly from the eigenvalues of
$\square$ because the relevant fields have a mode expansions purely in
terms of the scalar eigenfunctions on $T^{pq}$.  In \DimEn\ and
\DimEnTwo\ we have summarized the spectrum of operators which in
\cite{KRvN} corresponded to the first, third, and sixth Kaluza-Klein
towers of scalars.  To obtain the rest of the bosonic spectrum, one
must deal with the analogues of the vector, symmetric tensor, and
antisymmetric tensor spherical harmonics appearing in (2.20) of
\cite{KRvN}; for the fermionic spectrum, a study of the eigenfunctions
of the Dirac operator is also required.  We leave these more involved
studies for future work.

Inspection of the spectrum of dimensions on all the $T^{pq}$ reveals a
rather uninteresting sequence of numbers, mostly complicated
irrationals.  However, $T^{11}$ exhibits a fascinating feature which
finds its explanation in the superconformal algebra.  Operators of
algebraically protected dimension are typically those whose dimension
is the lowest possible given a certain $R$-charge, or else descendants
of such operators derived by a series of supersymmetry
transformations.  The $R$-symmetry in this context is the $U(1)$ part
of the isometry group of $T^{11}$, which acts by shifting $\alpha_3$.
The integer $R$-charge $k$ is related to $m_3$ by $m_3 = k/\sqrt{2}$.
Without loss of generality let us take $k \geq 0$.  Using \EForm,
\DimEn, and \DimEnTwo, one finds that the smallest possible value of
$\Delta$ is $\Delta = 3k/2$, and corresponds to a mode of
$h^\alpha_\alpha$ and $a_{\alpha\beta\gamma\delta}$ with $l_1 = l_2 =
k/2$ and $|m_1|,|m_2| \leq k/2$.  Thus we find a set of operators
filling out a $({\bf k+1},{\bf k+1})_k$ multiplet of $SU(2) \times
SU(2) \times U(1)$, where $({\bf d_1},{\bf d_2})_r$ indicates
$R$-charge $r$ and $SU(2)$ representations of dimensions $d_1$ and
$d_2$.  Indeed, $\Delta=3k/2$ saturates the algebraic bound on
$\Delta$ following from the superconformal algebra \cite{Sohnius}.
This part of the spectrum was anticipated on field theory grounds in
\cite{kw}, and it was argued there that the form of the operators
is $\tr (AB)^k$.  The related operators in the third and sixth towers
are descendants which presumably have the form $\tr F_1^2 (AB)^k + \tr
F_2^2 (BA)^k$ and $\tr F_1^4 (AB)^k + \tr F_2^4 (BA)^k$, where $F^4$
is the special Lorentz contraction $F_{\mu_1}{}^{\mu_2}
F_{\mu_2}{}^{\mu_3} F_{\mu_3}{}^{\mu_4} F_{\mu_4}{}^{\mu_1} -
\tf{1}{4} (F_{\mu_1}{}^{\mu_2}
F_{\mu_2}{}^{\mu_1})^2$.\footnote{Thanks to I.~Klebanov for a
discussion on this point.}

The supergravity predicts in addition a spectrum of operators with at
least one $SU(2)$ spin larger than the $R$-charge.  These operators
far outnumber the chiral primaries $\tr (AB)^k$: there are
$N_\chi(\Delta) \sim \tf{8}{81} \Delta^3$ such chiral primaries with
dimension less than $\Delta$, versus a number $N(\Delta) \sim
\tf{4}{405} \Delta^5$ of operators with larger $SU(2)$ spins, as
follows from Weyl's Law for the growth of eigenvalues of $\square\,$:
$N(\Delta) \sim {\Vol M_5 \over 60 \pi^3} \Delta^5$.  The dimensions
of these non-chiral operators are in general irrational (square roots
of integers).  However, there are special nonnegative integer values
of $n_1$ and $n_2$ such that for $l_1 = n_1+k/2$ and $l_2 = n_2+k/2$,
the dimensions of the operators in the $({\bf 2 l_1 + 1},{\bf 2 l_2 +
1})_k$ multiplet are again integer or half-integer: 
  $E = \left[ 2(n_1+n_2+1) + \tf{3}{2} k \right]^2 - 4$, so 
  $\Delta = 2(n_1+n_2+2) + \tf{3}{2} k$ 
 for relatives of the dilaton and 
  $\Delta = 2(n_1+n_2) + \tf{3}{2} k$ or 
  $2(n_1+n_2+4) + \tf{3}{2} k$
 for relatives of $h^\alpha_\alpha$ and $a_{\alpha\beta\gamma\delta}$.
The $n_i$ for which this occurs are solutions to the Diophantine
equation $n_1^2 + n_2^2 - 4 n_1 n_2 - n_1 - n_2 = 0$: that is,
consecutive terms in the sequence
$\{0,0,1,5,20,76,285,\ldots\}$.\footnote{Let us briefly indicate the
solution of the Diophantine equation.  Define
  $$
   f_{\pm}(n) = {1 + 4n \pm \sqrt{1 + 12 n + 12 n^2} \over 2} \ .
  $$
 By construction, any pair $(n_1,n_2) = (f_-(n),n)$ solves the
equation, and any solution must have this form up to interchange of
$n_1$ and $n_2$.  Also, $f_+^{(i)}(0)$ is always a nonnegative
integer, where the superscript denotes iterative application of $f+$.
To see that $(n_1,n_2) = (f_-(f_+^{(i)}(0)),f_+^{(i)}(0))$ where
$i=0,1,2,3,\ldots$ are the only solutions, note that $f_-(f_+(n)) = n$
for nonnegative $n$, and $f_+(f_-(n)) = n$ for positive $n$.
Furthermore, $f_-(n) \leq n$ for nonnegative integer $n$, with
equality iff $n = 0$; and $f_-(n)=0$ only for $n=0$ or $1$.  Starting
with any putative solution $(f_-(n),n)$, one can find successively
smaller solutions by repeated application of $f_-$ to $n$ until one
reaches $n=0$.  The properties of $f_\pm$ now guarantee that $n$ is a
member of the sequence $\{f_+^{(i)}(0)\}_{i=0}^\infty =
\{0,1,5,20,76,285,\ldots\}$.}

If these higher dimension operators are, in some exotic sense,
algebraic descendants of chiral primaries, then since the $SU(2)$
spins for given $R$-charge are larger than for the chiral primaries,
the relevant algebra must involve the $SU(2)$'s.  But, in contrast to
the $U(1)$ $R$-symmetry group, the $SU(2)$'s do not participate in the
usual ${\cal N}=1$ superconformal algebra.  Besides, the nilpotency of
supersymmetry generators implies that the dimension of a descendant
cannot be arbitrarily larger than the dimension of its chiral primary
parent.  So we do not see any reason why these peculiar series of
operators with integer or half-integer dimensions should be protected.

\section{Conclusions}
\label{Conclude}

It is an open question whether a quantum field theory constructed
holographically from compactified string theory must include a local
gauge invariance.  Supergravity, and by extension closed strings, see
only the gauge invariant observables.  One can argue that if the
compactification is supported by Ramond-Ramond charge, then it should
be realizable in string theory as a D-brane configuration, and gauge
invariance emerges from the dynamics of open strings attached to the
branes.  In light of this reasoning it would be interesting to find
singular six-manifolds to which the $T^{pq}$ spaces are related in the
same way $T^{11}$ is related to the conifold $z_1^2 + z_2^2 + z_3^2 +
z_4^2 = 0$ \cite{kw}.  An analysis of D-branes on such manifolds
should reveal a weak coupling gauge theory version of the conformal
field theories defined holographically in this paper.

Perhaps the most general statement that can be made about
supersymmetric theories constructed holographically from a product
space $AdS_5 \times M_5$ is that the $R$-current must be free of
gravitational anomalies: $\partial_\mu \langle T_{\alpha\beta}
T_{\gamma\delta} R^\mu \rangle = 0$.  This statement is nontrivial for
${\cal N} = 1$ and~$2$ supersymmetric theories, where the $R$-symmetry
is $U(1)$ or~$U(2)$; for ${\cal N}=4$ the $R$-symmetry group is
$SU(4)$, which is simple and hence automatically free of gravitational
anomalies.

The original correspondence \cite{JuanAdS97,gkPol,WitHolOne} between
$AdS_5 \times S^5$ and ${\cal N} = 4$ super-Yang Mills theory has been
questioned because so much of what it predicts is either
non-verifiable (for instance the coefficient on the $q\bar q$
potential \cite{jWilson,ReyYee}) or else largely a consequence of the
large supergroup apparent on both sides of the duality (for example
scaling dimensions of chiral operators).  Already, holography's
successes go beyond the constraints of symmetry in predicting Green's
functions \cite{MIT,lmrs} and in elucidating a geometric picture of
confinement \cite{WitHolTwo} and of baryons \cite{WitHolThree,go}.
Hopefully the example of D3-branes on conifolds, described for large
$N$ by the $T^{11}$ manifold, will serve as further evidence that
holography captures not only supergroup theory but gauge theory
dynamics.  The verification of $\gamma_{IR} = -1/2$ and $c_{IR}/c_{UV}
= 27/32$ seems particularly nontrivial, since both relations rely on
properties of the RG flow from the $S^5/{\bf Z}_2$ theory to the
$T^{11}$ theory.

\section*{Acknowledgements}

I would like to thank E.~Witten for correcting a crucial sign in
section~\ref{TpqVol} and for useful discussions on other points; and I
have benefitted from discussions from I.~Klebanov, N.~Seiberg, and
O.~Aharony.  This research was supported in part by the National
Science Foundation under Grant No. PHY94-07194, by the Department of
Energy under Grant No. DE-FG02-91ER40671, by the James S.~McDonnell
Foundation under Grant No. 91-48, and by the Hertz Foundation.

\newpage

\bibliography{squash}
\bibliographystyle{ssg}

\end{document}